\documentclass[conference, letterpaper]{IEEEtran}

\usepackage{algorithm}
\usepackage{algorithmic}
\usepackage{color}
\usepackage{subfigure}
\usepackage{epsfig}
\usepackage{setspace}
\usepackage{float}

\usepackage[cmex10]{amsmath}
\usepackage{color}
\usepackage{fancyhdr}
\usepackage[caption=false,font=footnotesize]{subfig}

\renewcommand{\thispagestyle}[2]{}

\fancypagestyle{plain}{
        \fancyhead{}
        \fancyhead[C]{first page center header}
        \fancyfoot{}
        \fancyfoot[C]{first page center footer}
}
\begin{document}

%
\title{The Art of Crypto Currencies \\ {\large A Comprehensive Analysis of Popular Crypto Currencies}}

\author{Sufian Hameed, Sameet Farooq\\
       IT Security Labs, National University of Computer and Emerging Sciences (FAST-NUCES), Pakistan\\
       sufian.hameed@nu.edu.pk}

\maketitle

\begin{abstract}
Crypto Currencies have recently gained enormous popularity amongst the general public. With each passing day, more and more companies are radically accepting crypto currencies in their payment systems, paving way for an economic revolution. Currently more than 700 crypto-currencies are available at Coindesk alone for trade purposes. As of November 2016, the Crypto currencies hold a total market share of over 14 Billion USD\footnote{The price and overall market capitalization might change at the time this article is being read.} \cite{marketcap}. With no centralized institution to monitor the movement of funds, Crypto currencies and their users are susceptible to multiple threats. In this paper we present an effort to explain the functionality of some of the most popular crypto currencies available in the online market. We present an analysis of the mining methodologies employed by these currencies to induce new currency into the market and how they compete with each other to provide fast, decentralized transactions to the users. We also share, some of the most dangerous attacks that can be placed on these crypto currencies and how the overall model of the crypto currencies mitigates these attacks. Towards the end, we will present taxonomy of the five highly popular crypto currencies and compare their features.
\end{abstract}


\begin{IEEEkeywords}
Crypto Currency; Bitcoin; Ripple; Litecoin; Dash coin; Stellar
\end{IEEEkeywords}

%
\IEEEpeerreviewmaketitle

\section{Introduction}

The need for crypto currency as laid down by Timothy May was to ensure the possibility of anonymous transactions of money and to create a society wherein secrecy and privacy are the prevailing features \cite{TMay}. This initial idea was capitalized by many researchers and activists of Cypherpunk to make a practical application of cryptography in all spheres of life.

Wei Dai came up with the outlining protocol for "B-Money" in 1998 that was practical in nature and was introduced as a bi-product of Timothy May's Crypto-Anarchy \cite{BMoney}. The B-Money laid out the foundation on which Satoshi 10 years later constructed Bitcoin \cite{nakamoto2008bitcoin}, the first decentralized crypto currency to be made publicly available.

The modern era is a digital era wherein the concept of fiat money has been challenged by crypto currencies. Crypto currencies are an alternative to fiat currencies with no central authority controlling the generation of money. The crypto currencies are different from conventional fiat system of currencies where no federal signatory governs the flow of currency. Unlike the fiat currency where the federal banks or governments are responsible for the generation and printing of money, the crypto currency is generated by a process called mining. Crypto currencies use complex hashing and time stamping methodologies to uniquely identify each coin within that currency \cite{cryptocurrency}. Crypto currency systems generally claim to provide anonymous, decentralized processing of transactions. This anonymity can be used as an additive preventive measure for user confidentiality and privacy.

The acceptance and demand of crypto currencies has increased a hundred fold over the past few years. Similarly, the industry around crypto currencies has evolved since its inception and a number of stake holders are now associated with the growing trade and acceptance of crypto currencies. Currently, crypto currencies are readily available at hundreds of exchanges around the world against fiat currency. Many large companies are now adopting crypto currencies into their payment systems. Bitcoins are easily accepted at Microsoft, Wordpress, Amazon, Apple’s App store, Wikipedia, Dell and other major brands in different sectors of life \cite{bitcoinusage}. However, if a store is unwilling to accept crypto currencies in their online payment systems, coins can be converted to physical world goods via gift cards. Many gift card businesses accept major crypto currencies such as Bitcoin and provide the customers with gift cards to be availed at a physical store \cite{bitcoinpurchase}.

\begin{figure*}
\centering
\epsfig{file=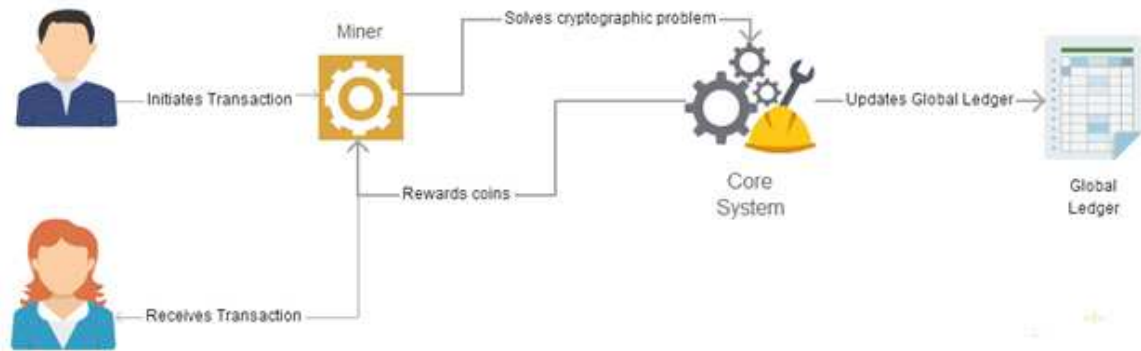, scale=0.7}
\caption{Architecture of Proof-of Work based Currency}
\label{fig:proof_of_work}
\end{figure*}

\begin{figure*}
\centering
\epsfig{file=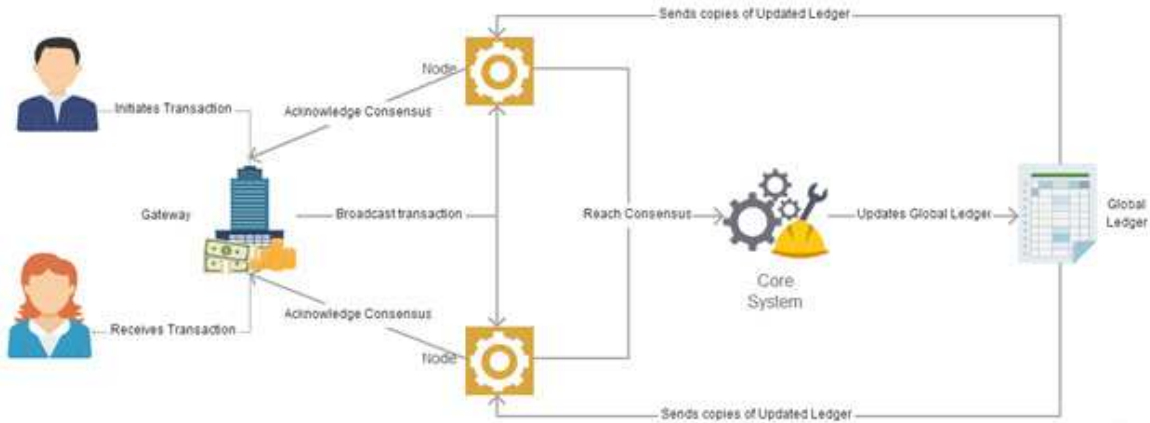, scale=0.7}
\caption{Architecture of Consensus Based Currency}
\label{fig:concensus-based}
\end{figure*}

The increasing requirement of computational power to solve mining problems has resulted in the need of mining pools and many of these pools currently exist for a new miner to join and generate new currency. Similarly, cloud vendors provide mining resources on rental basis to miners, which has opened a new avenue of stake holders associated with mining of crypto currencies. Likewise, hardware companies have now built and configured special hardware components solely for the purpose of solving the mining problems and this has elevated the variety of the people associated with the rising industry of crypto currency itself.

As of now, hundreds of crypto currencies currently exist in the online market for trade purposes. These currencies are as expensive as Skidoo, one unit of which is equivalent to USD 2,350 dollars, to as cheap as GCoin, which is available for as low as few USD 0.000000001 \cite{marketcap}. Crypto Currencies can generally be classified as either Proof-of Work based currency or Consensus Based Currencies, depending on how they settle transactions within their devised protocols. Figure. 1 and Figure. 2 depict the overall architecture of the two genres of crypto currencies.

\begin{itemize}
  \item \textbf{Proof-of-Work scheme:} This scheme add some work or difficulty to validate the transaction. In particular proof-of-work systems repeatedly run difficult hashing algorithms or other client puzzles to validate the electronic transactions.
  \item \textbf{Consensus scheme (Proof-of-Stake):}  This scheme aims to achieve distributed consensus by asking users to prove ownership or their stake in the currency.
\end{itemize}

In this paper we carry out a survey of the five most popular crypto currencies on the basis of their market capitalization and compare their working and performance with each other. For each of the currency, we will explain the working model of the currency, the mining approach through which new currency is generated in the market. We will also explain some of the eminent limitations in the protocols and mitigation actions taken by each protocol to overcome these limitations.

\section{Bitcoin}

\begin{figure*}
\centering
\epsfig{file=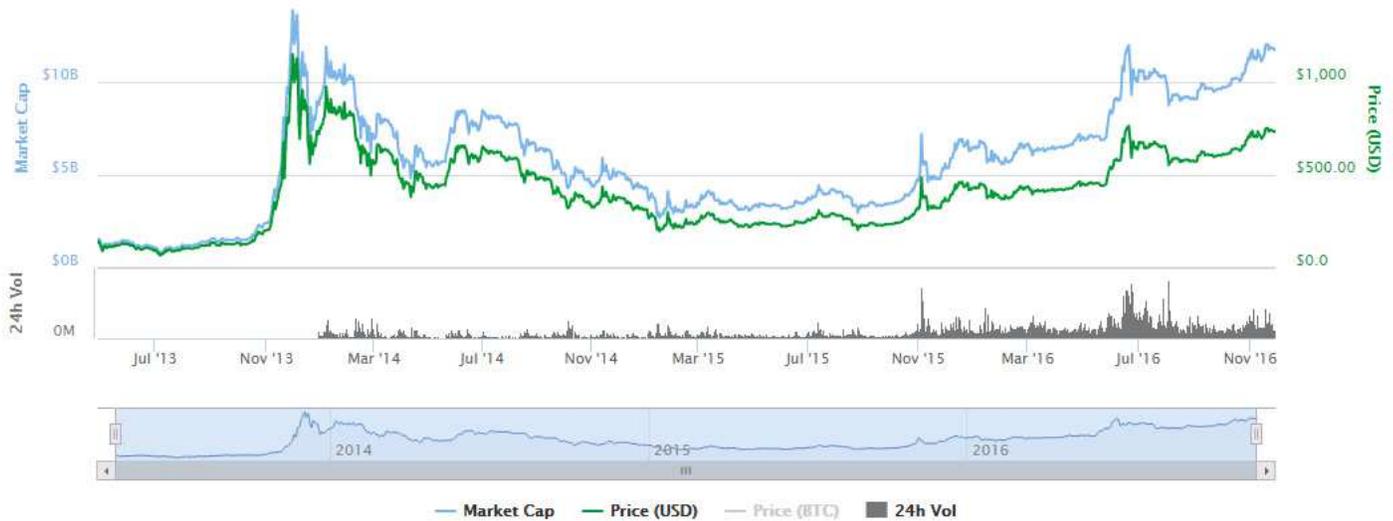,scale=0.5}
\caption{Price and Market Capitalization of Bitcoin \cite{bitcoinmarket}}
\label{fig:bitcoin_market}
\end{figure*}

Bitcoin is the first decentralized crypto currency established in 2008 based on the work of a pseudonymous developer and researcher Satoshi Nakomoto \cite{nakamoto2008bitcoin}. It is a peer to peer network of nodes in which all nodes maintain a single copy of transactions known as Ledger. Bitcoin was proposed as a solution to double spending attack. Bitcoin has emerged as a widely acceptable medium of transaction over the past few years. As of November 2016, the price of one Bitcoin is above USD 700 and the Bitcoin has a market capitalization of over 11.9 billion dollars making it the most popular crypto currency to date (see fig. \ref{fig:bitcoin_market} for overall price and market capitalization of Bitcoin).

\subsection{Design and Working}

\begin{figure}
\centering
\epsfig{file=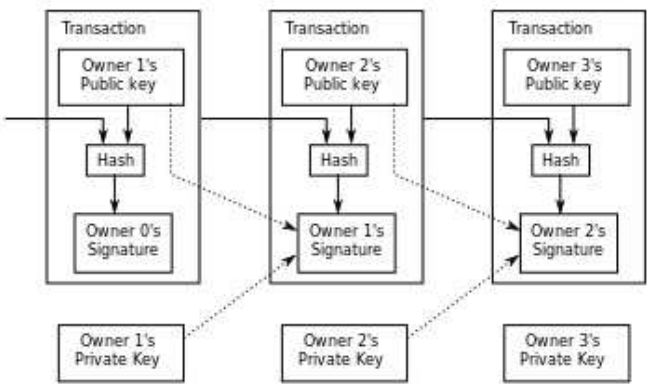,scale=0.7}
\caption{Merkle Hashing Process in Block Generation \cite{merkelhash}}
\label{fig:merkel_hash}
\end{figure}

In order to understand the architecture presented in Bitcoin, one has to familiarize with the following keywords that formulate the entire architecture. A coin is a chain of digitally signed certificates. A transaction is the process in which a sender of a coin digitally signs the hash of the previous transaction with the public key of the receiver and adds it to the end of the coin (see fig. \ref{fig:merkel_hash}). A timestamp server contains the timestamp of each transaction initiated within the system. The block is a collection of transactions that need to be validated and broadcasted by the nodes running the Bitcoin core engine.

Each block is a Merkle Root of the transaction, meaning that each block contains the hash of previous block and a nonce to satisfy the requirements of hashing function which in case of Bitcoin is SHA-256 (see fig. \ref{fig:merkel_hash}). When a transaction takes place, it is broadcasted to all the nodes within the network. The nodes form a block of transactions and work on finding the difficult proof-of-work for its block. The proof-of-work is the validation step in which nodes spend computational resources to find the correct nonce that fulfills the requirement of the hashing algorithm thereby validating the transactions in the block. Once a proof-of-work has been computed, it is broadcasted to all the nodes in the network and the block gets added into the block chain.

\subsection{Bitcoin Mining}

Since Bitcoin is a decentralized crypto currency, it implies that no regulatory body is in control of producing new coins. However, Bitcoins are added in the online market via a process called mining. Similar to the conventional meaning of the word where the labor usually works to find precious metals within the ground, mining is a mechanism wherein the nodes labor their way to solving the proof-of-work problem of each block chain validation. As a result of successful validation, the nodes get rewarded a sum of 50 Bitcoins. This is how the new Bitcoins are added into the system. Bitcoin block mining reward halves every 210,000 blocks to match the effect of hyperinflation. Current block mining reward is 12.5 bitcoins which is expected to half in approx 3.5 years based on current projections \cite{blockreward}.

\subsection{Possible Attacks and Mitigation Actions}

Over the years, the Bitcoin system has been under criticism due to its following vulnerabilities, many of which have been exploited so far.

\subsubsection{51 percent Attack}

This by far remains the greatest danger to the existence of Bitcoin and implies that a pool of dishonest nodes within the Bitcoin network gain 51 percent computational power of the entire network. If this scenario occurs, the security of the network is compromised since the controlling pool can take decisions on how the consensus is to be reached for each subsequent block. This allows the controlling pool to spend money that was not theirs to spend thereby inducing the "double spending problem". Only recently, GashIO \cite{gHash}, a Bitcoin mining pool, reached the 51\% computational mark thereby causing panic in the entire Bitcoin community \cite{51attack}.

The Bitcoin protocol was designed considering the fact that 51\% of the nodes will remain honest to the network. However the protocol is setup to pick random nodes for mining thereby breaking the power of computation and disallowing attackers to gain 51\% of attacking nodes in the system.

\subsubsection{Double Spending Problem}

A Bitcoin transaction usually takes 10 minutes before it is confirmed by the system. This waiting time is not acceptable to people who want transaction processing at a fast rate without waiting for confirmation. This gives rise to double spending problem without requiring 51\% percent computational power. Researchers have been able to successfully carry out the double spending attack by broadcasting fraudulent transactions to a large number of nodes along with honest transactions. This allows the network to assume that fraudulent transactions should be accepted by the network as they get accepted in most of the nodes \cite{bradbury2013problem}.

\subsubsection{Dust Transactions}

Another major limitation with Bitcoin was the increasing size of block chain. Previously the minimum amount for transaction was 1 Satoshi which caused block chain to reach upto 8GB in size \cite{8GB}. However the protocol was redesigned to change the minimum transaction amount of 5430 Satoshis which resulted in a smaller block chain.

\begin{figure*}
\centering
\epsfig{file=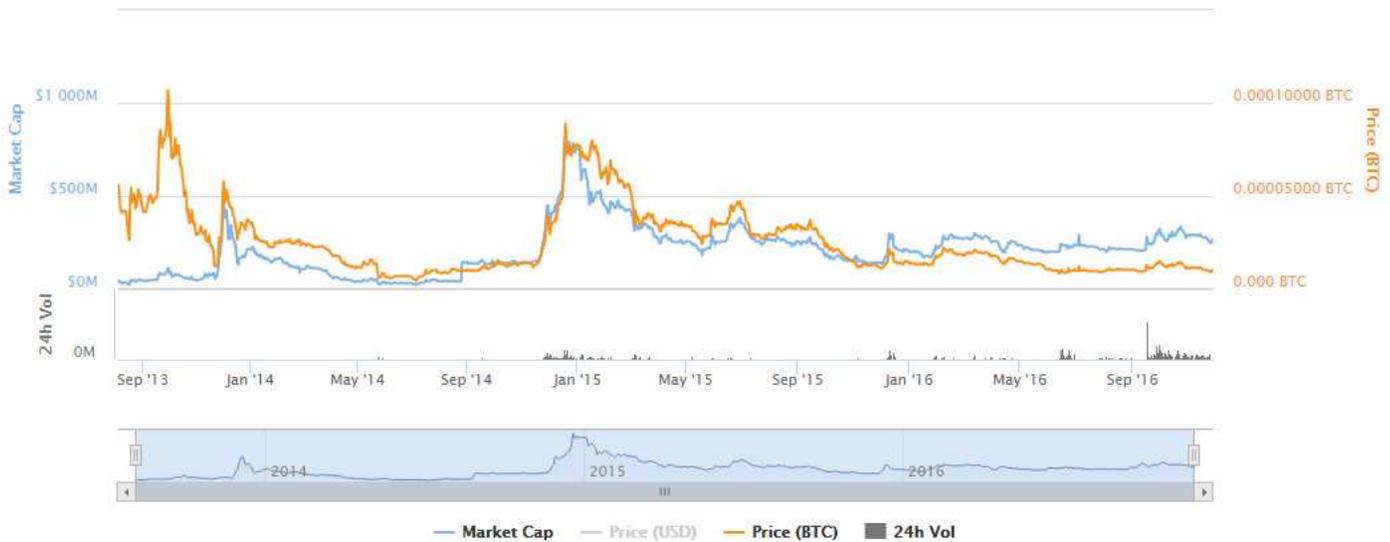,scale=0.5}
\caption{Price and Market Capitalization of Ripple \cite{ripplemarket}}
\label{fig:ripple_market}
\end{figure*}

\section{Ripple}

Ripple is the third most popular crypto currency as of today and has a market capitalization of over USD 249 million. Ripple was formulated by the Ripple Labs to support the electronic cash payment system known as Ripple Consensus Protocol \cite{schwartz2014ripple}. The Ripple consensus protocol was aimed at facilitating high speed transaction processing to financial institutions and individuals at minimal fee. Ripple (XRP) works within the Ripple Network as a bridging currency between different fiat currencies and unlike the fiat currencies, it is only worth what the other person is willing to pay for it. Currently 1 Ripple is priced at 0.00695 USD (see fig. \ref{fig:ripple_market} for overall price and market capitalization of Ripple).

\subsection{Design and Working}

Ripple Consensus Protocol Algorithm (RPCA) was proposed keeping in view the three major challenges of decentralized payment networks i.e. Correctness, Agreement and Utility and works on the principal of Byzantine Agreement \cite{schwartz2014ripple}. The RPCA consists of the following main components that we will be using throughout this section:

\begin{itemize}
  \item \textbf{Server:} An entity running the Ripple Server software and participating in the consensus process.
  \item \textbf{Ledger:} A complete record of the account balances of every user in network. A ledger is updated with each transaction successfully completing the consensus process.
  \item \textbf{Last-Closed Ledger:} The state of the ledger after completion of last successful consensus process. It also determines the current state of the entire network.
  \item \textbf{Open- Ledger:} A current state of the ledger which has not yet been confirmed by the network via the consensus process.
  \item \textbf{Unique Node List (UNL):} A subset of the server nodes that a particular server trusts. It is a list maintained by each Server and can be updated at any time.
  \item \textbf{Proposer:} A server broadcasting a set of transactions to be considered in the next consensus round.
\end{itemize}

Consensus process in RPCA takes place every 3 seconds and a Server 'S' takes all the valid transactions it has encountered between the last closed ledger period and the new consensus process and formulates a candidate set. This set is then broadcasted to all the UNLs of 'S'. All the UNLs are required to vote on the validity of the transactions based on the time stamp and balances that lay in their copy of the last closed ledger. If a minimum number of UNLs accept the transaction, the transaction is forwarded to the next round of consensus; else it is discarded to be taken into consideration in the next consensus process \cite{schwartz2014ripple}.

In the second round of consensus, the RPCA requires at least 80\% of the UNLs to agree on a valid transaction from first round. The requirement of 80\% of the UNLs is to ensure the correctness of the transaction and enforce that 80\% of the nodes have to be honest within the network for a transaction to be validated. In order to ensure the utility of the protocol, the consensus takes place every 3 seconds, allowing the users to send and receive currency at a very high speed.

Based on the RPCA, the ripple network uses Gateways as an entry point into the network. A client can create an account with the trusted Gateway to send currency to other untrusted clients. Ripple Network allows different currencies to be maintained in its system, allowing the clients to leverage the transaction processing in whichever currency they required. With Ripple (XPR) as the native currency of the system, the clients can send and receive ripples without having to trust the Gateways.

\subsection{Ripple Generation and Distribution}

The Ripple System consists of 100 Billion Ripples generated and fed into the ledger at the time of initiation. The system cannot generate any further currency due to protocol restrictions. The creators gifted 80 Billion out of 100 Billion Ripples to Ripple Labs which are given away to the end consumers through one of the following programs \cite{xrpdistribution}

\begin{itemize}
  \item \textbf{Users:} Following the Paypal’s marketing strategy; new users are awarded free Ripples upon joining the Ripple Network.
  \item \textbf{Developers:} The developers are encouraged to catch bugs in the open source software and provide patches to the already reported bugs and rewarded with Ripples.
  \item \textbf{Merchants:} Ripples are distributed to the merchants for the amount of transactions they bring into the system.
  \item \textbf{Gateways:} The creators are trying to incentivize the running of system by creating strategic partners and granting Ripples as bounties to its partners.
  \item \textbf{Market Makers:} Financial institutions and Forex agents are compensated specially for bringing liquidity to the model.
  \item \textbf{Administrative Cost:} The administrative cost of running the network, creating new products is all waged using Ripples.
\end{itemize}

\subsection{Possible Attacks and Mitigation Actions}

Ripple network has a number of advantages as it was developed to improve on Bitcoin itself. In this section, we will determine some of the most popular attacks in the crypto currency domain and how Ripple leverages its consensus protocol to overcome these threats.

\subsubsection{51 percent Attack}

The highly feared attack in the crypto currencies, the 51 percent attack, is being repelled by the Ripple Consensus Protocol by introducing UNLs. A server trusts transaction only its UNL trusts thereby constraining the attacker to get hold of nodes already in the UNL. This is highly non trivial as the resulting double spending transaction would infinitely regard the node as distrustful by the server. Also latency checks are placed to make sure all nodes are running effectively and mechanisms are in place to make sure that UNL gets updated if nodes show suspicious behavior.

\subsubsection{Denial of Service Attack:}

Ripple has mechanism in place for possible Denial of Service attack. For every transaction taking place in the system, 0.00001 XPR are destroyed by the system. Also a minimum of 20 XPR balance is to be maintained by a user in order to create a transaction in ledger \cite{xrpmm}. The idea behind the two is to bankrupt the attacker in case of a DoS attack by burning out Ripples and making transactions expensive. However, they will continue to be deemed cheap for average users.


\section{Litecoin}

\begin{figure*}
\centering
\epsfig{file=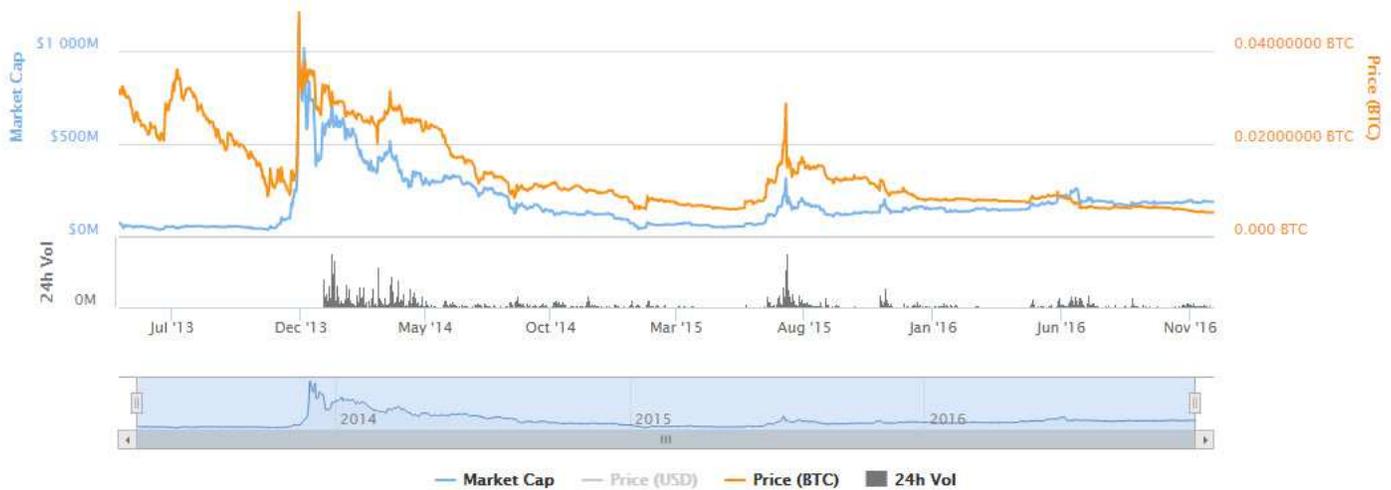,scale=0.5}
\caption{Price and Market Capitalization of Litecoin \cite{litecoinmarket}}
\label{fig:litecoin_market}
\end{figure*}

Created in 2011 by Charles Lee, a former Google engineer, the Litecoin was aimed at being "Silver to Bitcoin's Gold" \cite{litebit}. Litecoin is the fourth largest currency with an overall market capitalization of more than USD  189 million. At the time of writing, 1 Litecoin is available in the online market for USD 3.90 (see fig. \ref{fig:litecoin_market} for overall price and market capitalization of Litecoin). Litecoin was designed specifically to improve over the problems associated with Bitcoin and is the first crypto currency to have successfully deviated from the legacy proof-of-work of SHA-256 and implemented SCRYPT for block processing \cite{percival2009stronger}. This deviation led to multiple advantages of Litecoin, the details of which will be presented in the next sections.

\subsection{Design and Working}

Litecoin uses the same code base of Bitcoin with minor deviations in the protocol to improve on the gaps identified in Bitcoin. The aim of Litecoin was to provide faster transactions and reduce the impact of 51 percent attack. Litecoin blocks are generated at 2.5 minutes as compared to traditional 10 minute mark of Bitcoin. The faster block creation in Litecoin allows faster confirmation of transactions as blocks gets accepted into the block chain 4 times faster than Bitcoin. The working of Litecoin is quite similar to the working of Bitcoin which is explained in the previous sections. The key difference between the working of two crypto currencies lies in the difference of Proof-of-work mechanism explained in the Litecoin Mining section.

\subsection{Litecoin Mining}

Litecoin mining traditionally follows the same trajectory as Bitcoin mining with one major difference in the Proof-of-work mechanism. Bitcoin uses SHA-256 as its hashing algorithm in the Merkle Tree of the block chain, whereas Litecoin uses SCRYPT \cite{percival2009stronger} as its hashing algorithm. SCRYPT favors large amount of RAM memory and works in a serialized manner as compared to SHA-256 which is dependent on parallelization and computational power alone. This deviation of hashing algorithm led to the decentralization of Litecoin mining as it gives a chance to ordinary users with low computational power to participate in the mining process. Like Bitcoin, the Litecoin too rewards its mining nodes a sum 50 coins on solving a block. The sum is due to be halved every four years.

\subsection{Advantages of Litecoin}

Litecoin has a number of advantages over bitcoin which were highlighted in the previous sections but explained in this section:

\begin{itemize}
  \item SCRYPT mining is more feasible than SHA-256 mining. The reason for this is that SCRYPT uses fast access to large amounts of memory rather than depending on fast amount of arithmetic operations as required by SHA-256. With the development of ASICs (Application-Specific Integrated Circuit) for bitcoin mining, the modern computers and GPUs cannot participate in the Bitcoin Minining. However, ASICs are more expensive to design for SCRYPT as the device would require large amount of expensive RAM. This would allow modern GPUs and CPUs to participate in the mining process and get rewarded.
  \item Since the ASICs are not feasible for SCRYPT, it would decentralize the mining power and consequently limiting the dreaded 51\% attack. This would imply that no entity could be investing such a high amount of money to accumulate the mining power in order to carry out the double spending attack.
  \item Blocks are processed at 2.5 minutes rather than 10 minutes mark in Bitcoin. This allows faster confirmation of transaction. Also it allows more granularity in the network, e.g. a merchant can wait for two confirmations by the network and consume 5 minutes only as compared to one confirmation in Bitcoin that takes 10 minutes.
  \item The protocol implies that a total of 84 million Litecoins will be created in the life time, an amount four times greater than 21 million Bitcoins due to be created.
  \item The block retarget was 2016 in both Bitcoin and Litecoin. Since Litecoin blocks are generated 4 times faster, the difficulty mark needs to be adjusted every 3.5 days. The relatively quick adjustment in difficulty of the hashing function works fairly well in the event of a large number of miners suddenly dissipating from the network.
\end{itemize}	

\subsection{Limitations of Litecoin}

The above advantages of Litecoin have some limitations associated with themselves, some of which are discussed below.

\begin{itemize}
  \item In case of a Botnet attack, the owner could utilize the controlled bots for mining purposes and achieve a greater mining pool for itself. This would yield a higher benefit for the botnet owner as it would increase the probability of attacker solving the block chain problem. Where this attack is beneficial for the attacker, it also supplements the Litecoin network as the botnet controller brings with itself computational resources into the network. However the aim of crypto currency should be to put the greater good in front of individual benefits.
  \item The faster block generation results in bigger block chain. Accordingly, the block chain size would be 4x the block chain size of Bitcoin \cite{litebit}.
\end{itemize}


\section{Dash Coin}

\begin{figure*}
\centering
\epsfig{file=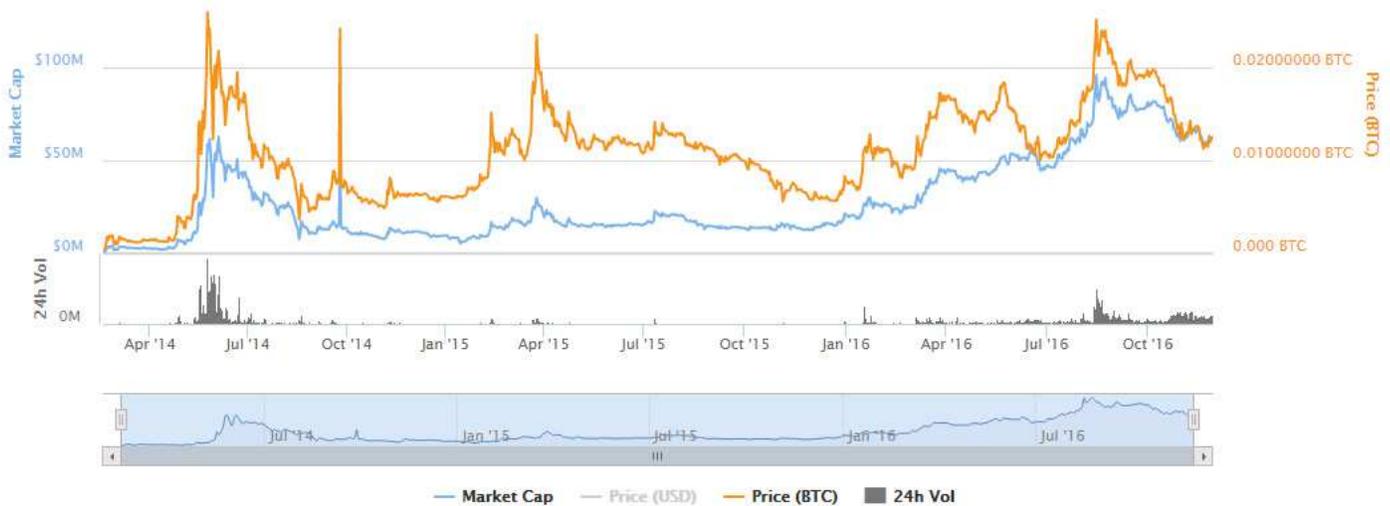,scale=0.5}
\caption{Price and Market Capitalization of Dash coin \cite{dashcoinmarket}}
\label{fig:dash_market}
\end{figure*}

Dash coin is another premier and one of the most popular crypto currencies, developed to add anonymity to Bitcoin \cite{duffield2014dash}. Dashcoin differs from Bitcoin in a number of ways. Firstly it is based on the countermeasure of CoinJoin \cite{maxwell2013coinjoin} to provide users of crypto currency with added security and anonymity over transactions. We will discuss the problem identified in CoinJoin in the following section. Secondly, it provides near instant transactions by employing a secondary network of Master Nodes. This implies that consensus must be reached within the quorums of Master Nodes on a transaction in order to accept the transactions. Currently, Dash coin is the seventh most popular currency in the list with a market capitalization of USD 62 Million. In the online market, Dash is available for trading at USD 9.10 per coin (see fig. \ref{fig:dash_market} for overall price and market capitalization of Dash coin).

\subsection{Design and Working}

Dash coin employs an incentivized secondary network of nodes known as Master Nodes to cater the problem of reducing mining nodes in Bitcoin. Within Dash, Master nodes are engaged to enhance the functionality of the entire network. Maser nodes are utilized for message sending, consensus as well as providing anonymity to the users. For all these functionalities, the Master nodes are rewarded 45\% of the total reward of block processing \cite{duffield2014dash}.

In order to deploy a Master Node, 1000 Dash must be reserved in the wallet as collateral. This amount is only reserved for a node to work as a Master node and receive its fair share of reward. If at any moment the limit for 1000 Dash is not fulfilled, the node gets ruled out of the Master Node list maintained by the client software. This puts an upper bound to the total number of Master Nodes running in the network. As the total available Dash is around 5.3 million, the total number of Master Nodes the system can accommodate is around 5300. Once a node has been categorized as Master Node, a ping message is sent every 15 minutes to check the liveliness of the node.

Master Nodes help in reaching consensus by the use of validating transaction locks. Once a transaction has been initiated, a transaction lock is propagated to all the Master Nodes in the network. If a consensus is reached on the transaction lock by a quorum of nodes, it gives rise to the probability that the transaction will be accepted by the block chain. The transaction thus gets accepted by the system on achieving consensus and all the subsequent conflicting transactions are rejected. If however a consensus is failed to reach, the transaction gets validated via conventional block processing. This is how Dash preserves the integrity of the system against double spending attack. This module is called InstantX due to the fact that it allows near instant transactions in the system as the time to reach consensus is fairly low as compared to Bitcoin \cite{duffieldtransaction}.

Dash also incorporates modules to preserve user's privacy. Although Bitcoin was developed to provide anonymous peer-to-peer transactions, academic research shows that it is possible for the observer to trace transactions back to the original user. This is done via forward linking or "through change linking" in which a user sends a proportion of total transaction amount to an identifiable source. The backward propagation from identifiable source results in loosing anonymity of the transaction as described by Gregory Maxwell, one of the core developers of Bitcoin, in his concept of CoinJoin \cite{maxwell2013coinjoin}. Dash overcomes these problems by Darksend, a module that uses the Master Nodes to merge/mix transactions of three different users into input and output sets so that at each round the input and the out values have equivalent sets with varying users. This reduces the probability of correctly observing a transaction to one-third. In order to complicate observation of transactions, a block chain approach is applied which elevates the complexity for the attacker.

\subsection{Dash Mining}

Dash uses X11 as its proof-of-work hashing instead of SHA-256 and SCRYPT, which are used by some of the other notable crypto currencies. The idea is to complicate the formation of ASICs and encourage traditional "hobbyist" mining. Dash is unique in the sense that it has a variable block reward that is based on difficulty. This means that while currently the block reward is 120, when difficulty rises the block reward will fall. Eventually the block reward will be driven down to its lowest amount which is 15 Dash. After that, every 2 years the block reward is halved again. So in 2 years, 7.5 Dash, in 4 years, 3.75 Dash, etc.

\subsection{Possible Attacks on Dash and Mitigation Actions}

In this section we consider some of the probable attacks on Dash Consensus Network and in the event of these attack how the system will prevent itself from colossal damage.

\subsection{Sybil Attack}

A probable Sybil attack would require the attacker to gain control of at least two-third of the entire Master Node network.  For a network of 1000 Master Nodes, the cost of adding 2000 further Master nodes to get 2/3 control of the network would require a demand of 2 million Dash. In a relatively small market of 5 million Dash, it would be difficult to get hold of 2 million Dash thereby thwarting the Sybil attack.

\subsection{Finney Attacks}

Finney Attacks implore that an attacker is mining a block normally. After the block has been processed, the attacker induces a transaction sending the payment back to him before another block has been attached to the block chain. This is prevented in the Dash by making sure locks are maintained by the Master Nodes until the transactions have been approved by consensus. The conflicting transactions are all rejected and discarded.

\begin{figure*}
\centering
\epsfig{file=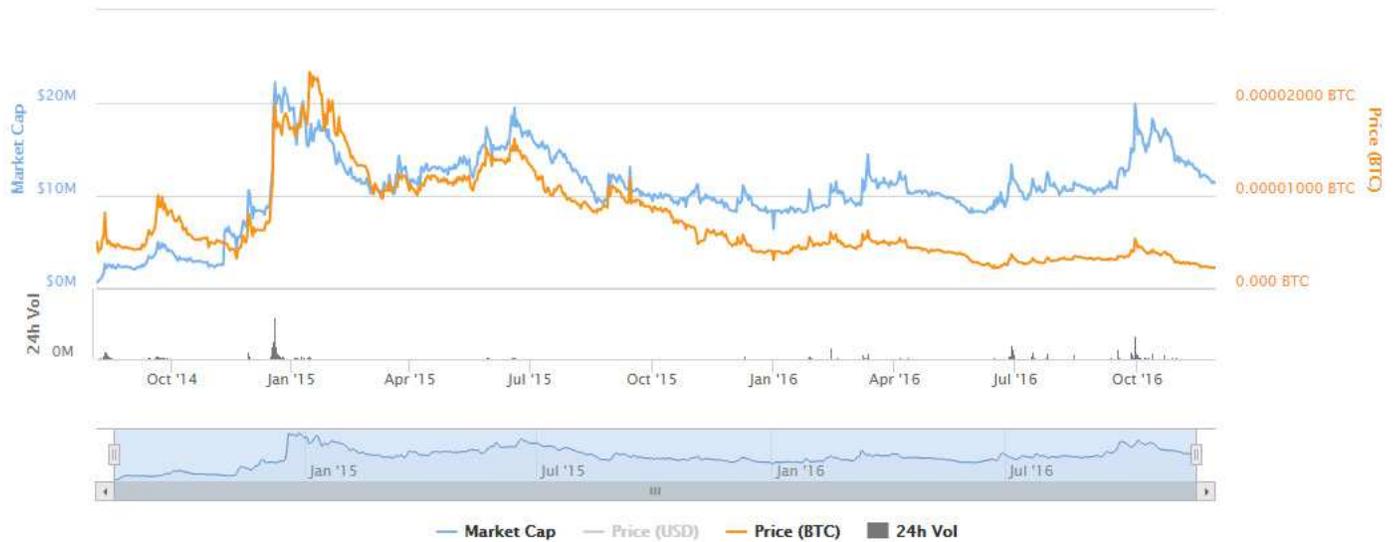,scale=0.5}
\caption{Price and Market Capitalization of Stellar \cite{stellarmarket}}
\label{fig:stellar_market}
\end{figure*}

\subsection{Transaction Lock Race}

The protocol can also be tested by transaction lock race in which an attacker submits two racing locks, one with payment to the merchant and one with payment back to him. In this scenario the network would be split between correct and attacking locks. It is the responsibility of the master nodes to chip in their vote and remove confusion from the network.

In case of other attacks, the conventional block processing methodology is used to validate the blocks there by securing the block chain.

\section{Stellar}

With the aim to provide the access to a greater number of people across the globe by lowering boundaries of membership into the system, Stellar is also one of the most popular cryptographic currencies having already surpassed the market capitalization of USD 11.5 million. Stellar is backed by a non-profit organization Stellar Development Foundation \cite{stellarfoundation} with an aim to spread financial literacy and access worldwide. Stellar, initially was based on Ripple however in May 2015, Stellar launched its unique consensus protocol, the Stellar Consensus Protocol \cite{mazieres2015stellar}, after limitations were identified in the former version. Most of the characteristics of the two currency exchange systems are the same. Currently, each Stellar is worth around USD 0.00169 (see fig. \ref{fig:stellar_market} for overall price and market capitalization of Stellar).

\subsection{Design and Working}

Stellar, like Ripple itself, acts as a bridging currency in the Stellar network. The difference between the two currency exchange networks is the consensus protocol each of them employs to validate a transaction. Stellar uses Federated Byzantine Agreement algorithm as a base around the Stellar Consensus Protocol (SCP). A transaction on a node in Stellar gets validated if a vast majority of the nodes it trusts validate the slot. The trusting nodes will in turn look to their own trusting nodes before calling any slot or transaction settled. The consensus is thus reached when a vast majority of the nodes in the network have called a transaction settled. This consensus is reached within a few seconds at most and results in fast processing of transactions. Stellar also claims to be the first protocol to provide decentralized control, flexible trust, low latency and asymptotic security all together.

An account on Stellar is identified by a unique "address", which is the (hashed) public key half of a public/private key pair in public-key cryptography. To spend the balance or change a property of an account in the ledger, the account holder must sign a corresponding "transaction" using the private key half of the account's key pair, and submit it to a Stellar server for propagation to the network. The Stellar server will check the authenticity of the digital signature to confirm the transaction is signed with the correct private key. A Stellar transaction is a signed instruction broadcast to the entire network which modifies the state of one or more accounts in the ledger. A set of transactions is applied to the ledger after a consensus round, and a new ledger is created. There are many different types of transactions that an account can create, including: Payment, OfferCreate, TrustSet, AccountSet.

\begin{table*}

\centering
\caption{Taxonomy of Crypto Currencies}

\begin{tabular}{|l|l|l|l|l|l|}
  \hline
  Currency & Scheme & Decentralize Control & Low Latency & Flexible Trust & Asymptotic Security\\
  \hline \hline
  Bitcoin & Proof-of-Work & Yes & No & No & No\\
  Ripple & Byzantine Agreement & No & Yes & Yes & Yes\\
  LiteCoin & Proof-of-Work & Yes & No & No & No\\
  Dash Coin & Proof-of-Work & Yes & Yes & Yes & Maybe\\
  Stellar & Federated Byzantine & Yes & Yes & Yes & Yes\\
   & Agreement &  &  &  & \\

  \hline
\end{tabular}

\label{taxonomy}
\end{table*}

\subsection{Stellar Generation and Distribution}

Stellars cannot be mined like other crypto currencies. At the time of starting, 100 billion Stellars were deposited in the "root" account. These stellars are due to be utilized in the following manner \cite{stellardistribution}.

\begin{itemize}
  \item \textbf{Signup Program:} Signing up with Stellar gets you free Stellars. Stellar uses Facebook for identifying spam or duplicate accounts. 50 Billion Stellars are to be distributed via this method.
  \item \textbf{Non- Profit Organizations:} Since the idea behind Stellar is to promote financial reach to deserving people with limited connectivity or resources, Stellar plans to utilize 25\% of its resources to be given away to non-profit organizations which plan to deliver these to remote areas.
  \item \textbf{Bitcoin Program:} Stellar has reserved 20\% of the 100 billion stellars for the Bitcoin program. The idea behind this is to promote Bitcoin users to use stellar instead of Bitcoins with their wallets getting compensated in stellar.
  \item \textbf{Administrative Cost:} The remaining 5 percent of the Stellars will be used for administrative cost.
\end{itemize}

Apart from the above distribution, SCP also has a mechanism for inducting new coins into market via inflation. Theoretically, 1\% of the total Stellars are to be created every year. Every account can sign another account for getting new coins, the votes will be based on the number of Stellars in the account itself. Say, Alice has 120 Stellars in her account and she wants to vote for Bob to get the newly created Stellars. 120 Stellars in Alice's account will act as 120 votes for Bob. 50 contestants with most votes associated with their accounts will be rewarded the newly created Stellars.

\subsection{Limitations of Stellar}

Stellar by far claims to be the most secure consensus protocol but it has a few limitations associated with itself as pointed out by David Mazieres in his white paper for Stellar \cite{mazieres2015stellar}. Firstly, the protocol has a mechanism for locking out blocking sets of nodes but it has no mechanism for unblocking them. Secondly, the quorum slices are configurable by the user which threatens the integrity of the system if being incorrectly configured. Thirdly, widely trusted nodes can leverage their position in the market to spoof transactions.

As the SCP protocol is relatively new, other limitations are yet to be explored.

\section{Taxonomy}

After having laid down the design details and limitations of the five most popular crypto currencies in the world today, we would like to provide taxonomy of these digital currencies by comparing them against one another in table \ref{taxonomy}.

The matrix in table \ref{taxonomy} amply scrutinizes some of the major archetypes for monitoring the effectiveness of these crypto currencies. This includes the scheme of device, transaction control system of the crypto currencies, latency of transactions within their system, robustness of the system to changing trust network and security measures in place to ensure smooth transaction processing. As it is evident from the matrix, not all high functioning crypto currencies provide asymptotic security to its consumers. The pioneering currencies including Bitcoin as well as LiteCoin which occupy over 85\% of total market share are highly susceptible to the hazardous 51\% attack. Stellar, being the latest amongst this lot ensure highly decentralized control of its general ledger, providing fast, anonymous transaction service, while Ripple compromised on Decentralization of the monetization of the currency. Dash, as per the analysis has the best Collateral based system in place when it comes to Proof-of-Concept solving schemes.

\section{Conclusion}

Having understood the finer details associated with each of the popular crypto currencies, we have come a long way from where Satoshi Nakomoto gave his idea of a peer-to-peer system but with an overall market capitalization of approximately 14 billion dollars, crypto currencies have miles to go before they can ultimately replace the fiat currency.






%

%
%
\bibliographystyle{plain}
\bibliography{crypto}


\end{document}